\documentclass[amsmath,amssymb,12pt,superscriptaddress,nofootinbib]{revtex4}

\usepackage[dvipdfmx]{graphicx}
\usepackage{bm}
\usepackage{color}
\usepackage{soul} 
\usepackage{caption} 

\sethlcolor{yellow}

\newcommand{\dd}{\mathrm{d}}

\newcommand{\del}{\partial}

\definecolor{DarkBlue}{rgb}{0,0,0.7} 

\definecolor{DarkRed}{rgb}{0.65,0,0} 

\definecolor{DarkGreen}{rgb}{0,0.6,0}

\begin{document}
\baselineskip5.5mm
\thispagestyle{empty}

{\baselineskip0pt
\leftline{\baselineskip14pt\sl\vbox to0pt{
               \hbox{\it Division of Particle and Astrophysical Science}
              \hbox{\it Nagoya University}
                             \vss}}
\rightline{\baselineskip16pt\rm\vbox to20pt{
\vss}}%
}

\author{Masato Tokutake}\email{tokutake@gravity.phys.nagoya-u.ac.jp}
\affiliation{ 
Gravity and Particle Cosmology Group,
Division of Particle and Astrophysical Science,
Graduate School of Science, Nagoya University, 
Nagoya 464-8602, Japan
}

\author{Chul-Moon~Yoo}\email{yoo@gravity.phys.nagoya-u.ac.jp}
\affiliation{
Gravity and Particle Cosmology Group,
Division of Particle and Astrophysical Science,
Graduate School of Science, Nagoya University, 
Nagoya 464-8602, Japan
}

\vskip2cm
\title{Inverse Construction of the $\Lambda$LTB Model \\ from a Distance-redshift Relation}

\begin{abstract}
\vskip0.5cm

Spherically symmetric dust universe models with a 
positive cosmological constant $\Lambda$, 
known as $\Lambda$-Lema\^itre-Tolman-Bondi($\Lambda$LTB) models, 
are considered. 
We report a method to construct the $\Lambda$LTB model 
from a given distance-redshift relation 
observed at the symmetry center. 
The spherical inhomogeneity is assumed to be composed of
growing modes. 
We derive a set of ordinary differential equations for 
three functions of the redshift, which specify the spherical inhomogeneity. 
Once a distance-redshift relation is given, 
with careful treatment of possible singular points, 
we can uniquely determine the model by solving the differential equations 
for each value of $\Lambda$. 
As a demonstration, we fix the distance-redshift relation as 
that of the flat $\Lambda$CDM model with 
$(\Omega^{\rm dis}_{\rm m0}, \Omega^{\rm dis}_{\rm \Lambda 0})=(0.3,0.7)$, 
where $\Omega^{\rm dis}_{\rm m0}$ and  $\Omega^{\rm dis}_{\rm \Lambda 0}$ 
are the normalized matter density and the cosmological constant, respectively. 
Then, we construct the $\Lambda$LTB model for several values of $\Omega_{\rm \Lambda 0}:=\Lambda/(3H_0^2)$, where $H_0$ is the present Hubble parameter 
observed at the symmetry center. 
We obtain void structure around the symmetry center 
for $\Omega_{\Lambda 0}<\Omega^{\rm dis}_{\Lambda 0}$. 
We show the relation between the ratio $\Omega_{\Lambda0}/\Omega^{\rm dis}_{\Lambda 0}$ 
and the amplitude of the inhomogeneity. 
\end{abstract}

\maketitle
\pagebreak
\section{Introduction}

The cosmological principle is one of the most fundamental principles 
in cosmology, and established as a useful and successful working hypothesis. 
However, it is still of interest to consider the largest amplitude of 
cosmological scale inhomogeneity which can be compatible with 
the latest observational data. 
In other words, observational tests of the cosmological principle 
with precision measurements may be interesting subjects 
in observational cosmology. 
Since the isotropy of our universe is strongly supported by 
the isotropy of the cosmic microwave background(CMB), in this paper, 
we focus on spherically symmetric universe models with 
an observer at the symmetry center.

Spherically symmetric inhomogeneous universe models have attracted much attention 
as alternative models to explain the apparent accelerated expansion of 
our universe without 
a cosmological constant\cite{Zehavi:1998gz,Celerier:1999hp,Tomita:1999qn}. 
After the compatibility with the CMB anisotropy 
was discussed in Ref.~\cite{Alnes:2005rw}, 
many observational constraints on the spherical inhomogeneity 
were discussed by using Type Ia supernovae data, CMB anisotropy, 
baryon acoustic oscillation and 
so on(see, e.g. Ref.~\cite{Redlich:2014gga} for a recent detailed analysis). 
Those analyses, especially constraints from the kinematic Sunyaev-Zeldovich effect\cite{GarciaBellido:2008gd}, revealed that 
the apparent accelerated expansion cannot be explained only by introducing 
spherical inhomogeneity without a cosmological constant if we assume 
the standard history of our universe before the photon last scattering. 
If we do not assume the inflationary paradigm and the standard thermal history 
before the photon last scattering, 
the model significantly loses its predictability and comparable observations. 
However, even such eccentric models would be tested 
by future precision observations of 
the late time expansion of 
our universe\cite{Uzan:2008qp,Yoo:2008su,Quartin:2009xr,Yoo:2010hi,Yagi:2011bt}.

In this paper, we consider 
spherically symmetric dust universe models with a 
positive cosmological constant $\Lambda$, 
known as $\Lambda$-Lema\^itre-Tolman-Bondi($\Lambda$LTB) models.%
\footnote{Although the LTB solution originally contains 
the cosmological constant as a parameter, we use 
the word "$\Lambda$LTB" throughout this paper to avoid misconceptions. }
When we consider the relation between 
spherical inhomogeneity in $\Lambda$LTB models and 
observables, one of useful strategies is the inverse construction of 
the model inhomogeneity starting from given observables. 
A pioneering work has been done by 
Mustapha, Hellaby and Ellis in Ref.~\cite{Mustapha:1998jb}, 
where the angular diameter distance and the redshift-space mass density 
were supposed as the observables. 
The approach proposed in Ref.~\cite{Mustapha:1998jb} has been successfully 
performed in 
Refs.~\cite{Lu:2007gr,McClure:2007hy,Celerier:2009sv,Kolb:2009hn,Dunsby:2010ts}
for specific situations. 
Another important approach was proposed 
by Iguchi, Nakamura and Nakao(INN)\cite{Iguchi:2001sq}, 
where 
the inhomogeneity is assumed to be composed of growing modes.
This assumption is often adopted to guarantee the compatibility 
with the inflationary paradigm. 
The INN approach has been solved 
in Ref.~\cite{Yoo:2008su} in the whole redshift range. 
Apart from these two approaches, there are also several related works on the 
inverse construction\cite{Celerier:1999hp,Chung:2006xh,Vanderveld:2006rb,Romano:2009mr,Romano:2009ej,Krasinski:2014yja,Romano:2013bxa}. 

In this paper, along the INN approach, we explicitly construct the $\Lambda$LTB models 
whose distance-redshift relation coincides with that in the flat $\Lambda$CDM model 
with $(\Omega^{\rm dis}_{\rm m0}, \Omega^{\rm dis}_{\rm \Lambda 0})=(0.3,0.7)$, 
where $\Omega^{\rm dis}_{\rm m0}$ and  $\Omega^{\rm dis}_{\rm \Lambda 0}$ 
are the normalized matter density and the cosmological constant 
for the flat $\Lambda$CDM model, respectively. 
It should be noted that 
$\Omega^{\rm dis}_{\rm m0}$ and  $\Omega^{\rm dis}_{\rm \Lambda 0}$
are the parameters characterizing the distance-redshift relation, 
and $\Omega^{\rm dis}_{\rm \Lambda 0}$ is not necessarily 
equal to $\Omega_{\Lambda0}:=\Lambda/(3H_0^2)$, 
where $H_0$ is the present Hubble parameter observed at the symmetry center. 
Since we add a parameter $\Lambda$ to the case in 
the preceding works\cite{Iguchi:2001sq,Yoo:2008su,Yoo:2010qn}, 
we obtain an one-parameter family of solutions 
for a given set of values
$(\Omega^{\rm dis}_{\rm m0}, \Omega^{\rm dis}_{\rm \Lambda 0})$. 
The one-parameter family can be characterized by 
the parameter $\Omega_{\Lambda0}$. 
The difference between $\Omega_{\Lambda0}$ and $\Omega^{\rm dis}_{\Lambda0}$ 
can be regarded as a systematic error in estimation of $\Lambda$ 
due to the spherical inhomogeneity as is discussed in Refs.~\cite{Romano:2011mx,Negishi:2015oga}. 
In order to 
estimate the magnitude of the systematic error 
due to possible inhomogeneity, 
we calculate the amplitude of the inhomogeneity for several 
values of $\Omega_{\Lambda0}/\Omega^{\rm dis}_{\Lambda0}$. 
The method used in this paper is similar to that in Appendix of Ref.~\cite{Yoo:2010qn}, 
where a cosmological constant is not considered and 
the LTB solution can be described by a simple parametric form. 
In this paper, we work through all the complexity associated with  
a positive finite value of the cosmological constant(see Ref.~\cite{Valkenburg:2011tm} 
for fast accurate evaluation of metric components). 
Similar analysis has been done in Ref.~\cite{Negishi:2015oga} for perturbations 
on a homogeneous background.

In this paper, we use the geometrized units in which 
the speed of light and Newton's gravitational constant are one, 
respectively.

\section{Conditions to determine the $\Lambda$LTB model}
\label{set and eq}

\subsection{$\Lambda$LTB model and the radial geodesic}

We consider the $\Lambda$-Lema\^itre-Tolman-Bondi ($\Lambda$LTB) solution, 
whose line element is given by 
\begin{equation}
\dd s^2 = -\dd t^2+\frac{(\partial_r R(t,r))^2}{1-k(r)r^2}\dd r^2 
+ R^2(t,r)\dd \Omega^2, 
\end{equation}
where $k(r)$ is an arbitrary function of $r$ and $R(t,r)$ is the areal radius. 
The energy-momentum tensor is given by that of dust fluids: 
\begin{equation}
T^{\mu\nu} = \rho (t,r)u^\mu u^\nu, 
\end{equation}
where $\rho(t,r)$ is the mass density and $u^\mu$ is 
the 4-velocity of dust particles. 
From the Einstein equations, we obtain the following equation: 
\begin{equation}
(\partial_tR)^2 = -k(r)r^2+\frac{2M(r)}{R}+\frac{1}{3} \Lambda R^2, 
\label{Eeq}
\end{equation}
where $M(r)$ is an arbitrary function of $r$.  
By using $M(r)$, we can write $\rho(t,r)$ as follows:
\begin{equation}
\rho(t,r) = \frac{1}{4\pi}\frac{\partial_rM(r)}{R^2\partial_rR}. \label{densdef}
\end{equation}
For convenience, we introduce the following functions: 
\begin{equation}
m(r) := \frac{6M(r)}{r^3}\quad,\quad S(t,r) := \frac{R(t,r)}{r}. 
\end{equation}
Then, Eq.~\eqref{Eeq} is written as follows: 
\begin{equation}
(\partial_tS)^2 = f(r,S) := -k(r) + \frac{m(r)}{3S} + \frac{1}{3}\Lambda S^2.   
\label{Eeq2}
\end{equation}
Eq.~\eqref{Eeq2} can be integrated as 
\begin{equation}
t-t_B(r) = \int_{0}^{S}\frac{dX}{\sqrt{f(r,X)}} 
\label{Eeqsol}
\end{equation}
with an arbitrary function $t_B(r)$, 
which is called the bang time function because the areal radius 
$R$ vanishes at $t=t_B(r)$. 
When we consider the case $k(r)=$const. and $m(r)=$const., 
small perturbative inhomogeneity associated with $t_{\rm B}(r)$
is given by purely decaying modes in terms of cosmological perturbation theory. 
Therefore, in this paper, we simply assume 
that the bang time function is constant 
to guarantee the compatibility 
with the inflationary paradigm. 
The constant value of $t_{\rm B}$ can be eliminated by 
the time shift degree of freedom, namely, we can set $t_B(r) = 0$.

We assume that the observer is at the 
symmetry center $r = 0$. 
Then, we consider the past light-cone emanated from the symmetry center 
expressed by a trajectory parametrized by the redshift as follows: 
\begin{eqnarray}
t=t_{\rm lc}(z),  \\
r=r_{\rm lc}(z). 
\end{eqnarray}
Hereafter, for notational simplicity, 
we often omit the subscript ``lc".  
The null geodesic equations in the $\Lambda$LTB solution is given as
\begin{eqnarray}
\frac{\dd r}{\dd z} &=& 
\frac{1}{1+z}\frac{\sqrt{1-kr^2}}{\sqrt{f} + r\partial_t\partial_rS},  \label{Geq1} \\
\frac{\dd t}{\dd z} &=& 
-\frac{1}{1+z}\frac{S + r\partial_rS}{\sqrt{f} + r\partial_t\partial_rS}. \label{Geq2} 
\end{eqnarray}

\subsection{Conditions to determine the arbitrary functions}
The $\Lambda$LTB solution has three arbitrary functions $k(r)$, $m(r)$ and $t_B(r)$. 
One of these 
functional degrees of freedom 
corresponds to the gauge degree of freedom 
associated with the choice of the radial coordinate $r$. 
We impose a gauge condition for the radial coordinate $r$ and 
require the distance-redshift relation coincides with a given 
function $D_{\rm A}(z)$. 

We fix the gauge by imposing the light-cone gauge condition given by 
\begin{equation}
t(z)=t_0-r(z), 
\end{equation}
where $t_0$ is the present time at the central observer. 
From this condition, $t_{\rm lc}$ can be trivially given by $r_{\rm lc}$. 
Therefore, the remaining independent functions are 
$r_{\rm lc}(z)$, $k(r_{\rm lc}(z))$ and $m(r_{\rm lc}(z))$. 
Combining the light-cone gauge condition and the geodesic equations, we obtain 
\begin{equation}
\left.\left(r\partial_rS +S-\sqrt{1-kr^2}\right)\right|_{r=r_{\rm lc},~t=t_{\rm lc}}=0.  
\label{NC}
\end{equation}
We consider Eqs.~\eqref{Geq1}, \eqref{Geq2} and \eqref{NC} as independent equations.

The angular diameter distance on the past light-cone is given by 
$R(t(z),r(z))$. 
We impose that the angular diameter distance 
coincides with a given function $D_{\rm A}(z)$. 
In practice, we impose the following differential equation:
\begin{equation}
\frac{\dd R}{\dd z} = \frac{\dd D_A}{\dd z}. 
\label{DR}
\end{equation}
In this paper, for demonstration,  
we use the distance-redshift relation in a flat $\Lambda$CDM
universe instead of actual observational data. 
That is, we use the distance characterized by the cosmological 
parameters for the flat $\Lambda$CDM universe as follows: 
\begin{equation}
D_A(z) = D_{\rm \Lambda CDM}(z ; \Omega_{\rm m0}^{\rm{dis}},\Omega_{\Lambda0}^{\rm{dis}}), \label{dzrela}
\end{equation}
where $\Omega_{\rm m0}^{\rm{dis}}$ and $\Omega_{\Lambda 0}^{\rm{dis}}$ 
are the normalized matter density and the cosmological constant for 
the flat $\Lambda$CDM model. 
It should be noted that 
in a spherically symmetric inhomogeneous universe model, 
the Hubble parameter cannot be uniquely determined in 
off-center regions because of the difference between 
the radial direction and the transverse direction. 
The Hubble parameter can be uniquely defined only 
at the symmetry center. 
We define the present Hubble parameter $H_0$ as follows:
\begin{equation}
H_0:=\left(\del_t R/R\right)_{z=0}. 
\end{equation}
The normalization of the cosmological parameters, e.g. 
$\Omega_{\rm m0}^{\rm{dis}}$, is performed by using 
$H_0$. 
In our numerical calculations, we use the unit system given by $H_0=1$, 
and all dimensionful variables are normalized by $H_0$. 

\section{Derivation of differential equations}
\label{derive differential eq}

Let us derive the differential equations to determine three arbitrary functions
$r(z)$, $k(z)$ and $m(z)$.
Differentiating Eq.~\eqref{Eeq} with respect to $r$, we obtain 
\begin{eqnarray}
\partial_t\partial_rS 
&=& \frac{1}{2}f^{-1/2}\cr
&&\times
\left(-\partial_rk + \frac{S\partial_rm - m\partial_rS}{3S^2} + \frac{2}{3}\Lambda S\partial_rS\right).~~~~~
\end{eqnarray}
Multiplying $\dd r/\dd z$ to the above equation and using 
the null geodesic equations \eqref{Geq1} and \eqref{Geq2}, 
we get the following differential equation:
\begin{eqnarray}
&&\biggr[ \biggr(-\frac{m}{3S^2}+\frac{2}{3}\Lambda S\biggr)\biggr(\sqrt{1-kr^2}-S\biggr) + 2f\biggr] \frac{\dd r}{\dd z} \cr
&&- r\frac{\dd k}{\dd z} + \frac{r}{3S}\frac{\dd m}{\dd z} - \frac{2\sqrt{f}\sqrt{1-kr^2}}{1+z} = 0.
\label{ddEq1}
\end{eqnarray}

Differentiating Eq.~\eqref{Eeqsol} with respect to $r$, we obtain 
\begin{eqnarray}
0 
  & = & \frac{\partial_rS}{\sqrt{f(r,S)}} \cr 
  &&- \frac{1}{2}\int_{0}^{S}f(r,X)^{-2/3}\left(-\partial_rk + \frac{\partial_rm}{3X}\right)\dd X.~~\label{eq:drS}
\end{eqnarray}
Multiplying $\dd r/\dd z$ to the above equation and using Eq.~\eqref{NC}, 
we get the following differential equation: 
\begin{equation}
0 = \frac{\sqrt{1-kr^2}-S}{r\sqrt{f}}\frac{\dd r}{\dd z} - P\frac{\dd k}{\dd z} + \frac{Q}{3}\frac{\dd m}{\dd z},
\label{ddEq2}
\end{equation}
where
\begin{eqnarray}
P := -\frac{1}{2}\int_{0}^{S}f(r,X)^{-3/2}\dd X ,\\
Q := -\frac{1}{2}\int_{0}^{S}\frac{f(r,X)^{-3/2}}{X}\dd X.
\end{eqnarray}
These integrals can be numerically evaluated. 

Since, $\dd R/\dd z$ is calculated as
\begin{eqnarray}
\frac{\dd R}{\dd z} &=& \partial_tR\frac{\dd t}{\dd z} + \partial_rR\frac{\dd r}{\dd z} 
\cr
&=& \biggr(-r\sqrt{f} + \sqrt{1-kr^2}\biggr) \frac{\dd r}{\dd z}, 
\end{eqnarray}
from Eq.~\eqref{DR}, we get the following differential equation: 
\begin{equation}
\frac{\dd r}{\dd z} = \frac{1}{-r\sqrt{f} + \sqrt{1-kr^2}}\frac{\dd D_A(z)}{\dd z}.
\label{ddEq3}
\end{equation}

Using Eqs.~\eqref{ddEq1}, \eqref{ddEq2} and \eqref{ddEq3}, 
we can derive a set of three differential equations for $r(z)$, $k(z)$, $m(z)$
as follows: 
\begin{eqnarray}
\frac{\dd r}{\dd z} &=& A(r,k,m)\frac{\dd D_A(z)}{\dd z}, \label{dEq1}\\
\frac{\dd k}{\dd z} &=& B(r,k,m)\frac{\dd r}{\dd z} + \frac{1}{3S}\frac{\dd m}{\dd z} + C(r,k,m), \label{dEq2}\\
\frac{\dd m}{\dd z} &=& D(r,k,m)\frac{\dd r}{\dd z} + \frac{3P}{Q}\frac{\dd k}{\dd z}, \label{dEq3}
\end{eqnarray}
where
\begin{eqnarray}
A(r,k,m) &=& \frac{1}{-r\sqrt{f} + \sqrt{1-kr^2}}, \\
B(r,k,m) &=& \frac{1}{r}\biggr[ \biggr( -\frac{m}{3S^2} + \frac{2}{3}\Lambda S\biggr) 
\cr 
&&~~\times\biggr( \sqrt{1-kr^2}-S\biggr) + 2f\biggr], \\
C(r,k,m) &=& -\frac{2\sqrt{f}\sqrt{1-kr^2}}{r(1+z)}, \\
D(r,k,m) &=& -\frac{3}{rQ\sqrt{f}}\biggr( \sqrt{1-kr^2}-S\biggr).
\end{eqnarray}

\section{Regularity conditions and the solving method}
\label{boundary condition}
In the differential equations \eqref{dEq1}, \eqref{dEq2} and \eqref{dEq3}, 
there are two possible singular points(see also, e.g. Refs.~\cite{Krasinski:2014zza,Sundell:2016uqj}). 
One is at the center and the other is 
associated with the so-called critical point satisfying $\dd D_{\rm A}/\dd z=0$. 
At these points, we impose regularity conditions. 

\subsection{Regularity at the center}
We expand $r(z)$, $k(z)$ and $m(z)$ near the center as follows: 
\begin{eqnarray}
r(z) &=&  r_1z + \frac{1}{2}r_2z^2 +\mathcal{O}(z^3), \label{Taylor1} \\
k(z) &=& k_0 + k_1z  + \mathcal{O}(z^2), \label{Taylor2} \\
m(z) &=& m_0 + m_1z + \mathcal{O}(z^2), \label{Taylor3}
\end{eqnarray}
where we have assumed $r=0$ at $z=0$.
The right-hand side of Eq.~\eqref{dEq2} has the following term of the order $z^{-1}$: 
\begin{equation}
\sqrt{-k_0 + \frac{m_0}{3} + \frac{\Lambda}{3}}
\left(\sqrt{-k_0 + \frac{m_0}{3} + \frac{\Lambda}{3}} - H_0\right)z^{-1}.
\label{z-1}
\end{equation}
For the regularity at the center, 
we require the coefficient of this term vanishes at the center. 
Then, we obtain the following condition:
\begin{equation}
-k_0 + \frac{1}{3}m_0 + \frac{1}{3}\Lambda = H_0^2. 
\label{constraint at the center}
\end{equation}
This condition is consistent with the definition of $H_0$. 
Therefore we have a constraint for the three parameters 
$k_0/H_0^2$, $m_0/H_0^2$ and $\Lambda/H_0^2$. 
Hereafter, for convenience, let us consider $m_0/H_0^2$ and $\Lambda/H_0^2$ as 
the only independent parameters.

\subsection{Regularity at the critical point}
At the point satisfying $\dd D_{\rm A}/\dd z=0$, from Eq.~\eqref{dEq1}, 
we obtain $\dd r/\dd z=0$ unless $1/A(r,k,m) = 0$. 
The point with $\dd r/\dd z=0$ 
causes a unphysical solution with divergent physical quantities in general. 
Therefore, we impose $1/A(r,k,m) = 0$ at $z=z_{\rm{cr}}$ so that 
$r(z)$ can be a monotone increasing function of $z$. 
Let us consider the Taylor expansion near the critical point as follows:
\begin{eqnarray}
r(z) &=& r_{\rm{cr}} + r_{\rm{cr1}}(z-z_{\rm{cr}}) + \mathcal{O}((z-z_{\rm{cr}})^2), \\
k(z) &=& k_{\rm{cr}} + k_{\rm{cr1}}(z-z_{\rm{cr}}) + \mathcal{O}((z-z_{\rm{cr}})^2), \\
m(z) &=& m_{\rm{cr}} + m_{\rm{cr1}}(z-z_{\rm{cr}}) + \mathcal{O}((z-z_{\rm{cr}})^2). 
\end{eqnarray}
From the equation  $1/A(r_{\rm{cr}},k_{\rm{cr}},m_{\rm{cr}}) = 0$, 
we obtain the following equation:
\begin{equation}
m_{\rm{cr}}{r_{\rm{cr}}}^3 = 3D_A(z_{\rm{cr}}) - \Lambda {D_A(z_{\rm{cr}})}^3.
\label{constraint at the critical point}
\end{equation}
Since we can eliminate $r_{\rm{cr}}$ by using 
Eq.~\eqref{constraint at the critical point}, 
we consider $k_{\rm{cr}}$ and $m_{\rm{cr}}$ are the only independent parameters 
associated with the critical point.

\subsection{Newton-Raphson method}
As is shown in the previous subsections, independent parameters are $m_0/H_0^2$, 
$k_{\rm{cr}}/H_0^2$ and $m_{\rm{cr}}/H_0^2$ for a fixed value of $\Lambda/H_0^2$. 
To determine these parameters, we adopt the following method. 
First, we solve the differential equations from the center 
to a middle point $z=z_{\rm m}<z_{\rm cr}$
using a trial value of $m_0/H_0^2$. 
Second, we solve the differential equations from 
the critical point to the middle point 
using trial values of $k_{\rm{cr}}/H_0^2$ and $m_{\rm{cr}}/H_0^2$. 
Finally, we impose the smoothness conditions 
at the middle point $z=z_{\rm{m}}$ as follows:
\begin{eqnarray}
r_{\rm{m}-0}(m_0) - r_{\rm{m}+0}(k_{\rm{cr}},m_{\rm{cr}}) = 0, \nonumber\\
k_{\rm{m}-0}(m_0) - k_{\rm{m}+0}(k_{\rm{cr}},m_{\rm{cr}}) = 0, \label{NR}\\
m_{\rm{m}-0}(m_0) - m_{\rm{m}+0}(k_{\rm{cr}},m_{\rm{cr}}) = 0, \nonumber
\end{eqnarray}
where $X_{\rm{m}-0}$ and $X_{\rm{m}+0}$ are the values of 
$X$ at $z=z_{\rm{m}}$ when we solve from the center 
and the critical point, respectively. 
These smoothness conditions can be regarded as three independent conditions 
for $m_0/H_0^2$, $k_{\rm cr}/H_0^2$ and $m_{\rm cr}/H_0^2$. 
Then, we search for the values of $m_0/H_0^2$, $k_{\rm cr}/H_0^2$ and $m_{\rm cr}/H_0^2$
by using the 3-dimensional Newton-Raphson method. 
After the convergence, the smoothness conditions are satisfied within 
the accuracy $\sim10^{-10}$ in our numerical calculations. 
Eventually, we can obtain a unique solution for each set of 
a value of $\Lambda$ and a distance-redshift relation $D_{\rm A}(z)$.

\section{Solutions and density profile}
\label{density profile}
\begin{figure}[htbp]
\includegraphics[scale=0.6]{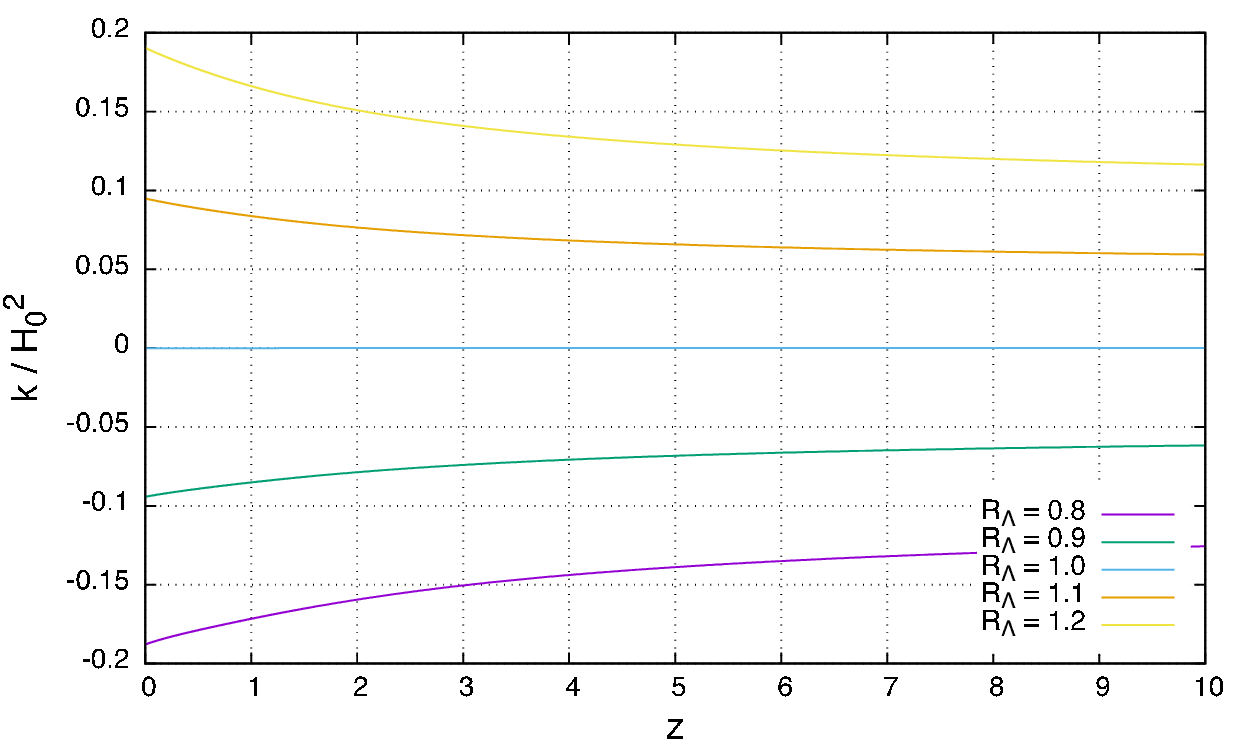}
\includegraphics[scale=0.6]{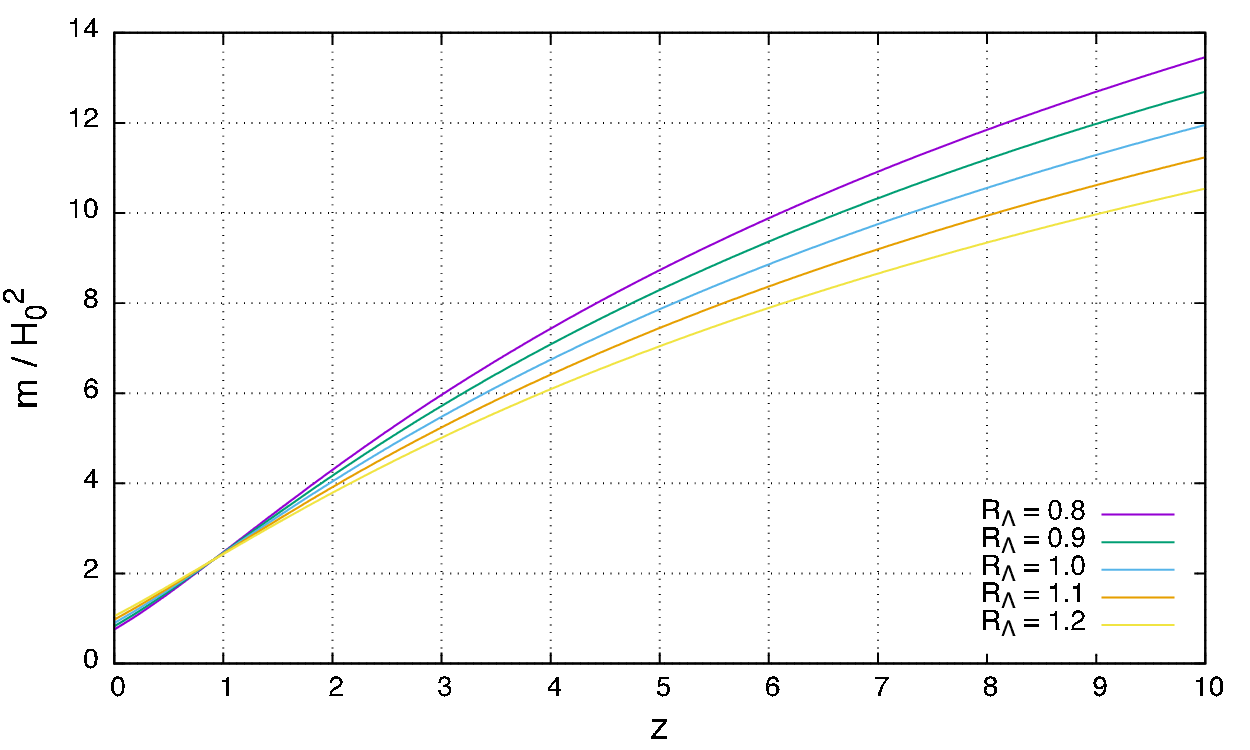}
			\caption{$k(r(z))/H_0^2$(left) and $m(r(z))/H_0^2$(right) are depicted as functions of $z$ for several values of $\mathcal R_\Lambda$. }
			\label{kmz}
\end{figure}

In this section, as a demonstration, 
we consider the case 
$D_A(z)
= D_{\rm \Lambda CDM}(z;0.3,0.7)$. 
We define $\mathcal R_\Lambda$ as 
\begin{equation}
\mathcal R_\Lambda := \frac{\Omega_{\Lambda 0}}{\Omega_{\Lambda 0}^{\rm{dis}}}, 
\end{equation}
where $\Omega_{\Lambda 0}:=\Lambda/(3H_0^2)$. 
We show $m(r(z))/H_0^2$ and $k(r(z))/H_0^2$ as functions of $z$ 
for several values of $\mathcal R_\Lambda$ in Fig.~\ref{kmz}. 

In these $\Lambda$LTB models, we can evaluate 
the density distribution on the present time slice $t=t_0$, 
where $t_0$ can be evaluated by 
\begin{equation}
t_0 = \int^{1}_{0} \frac{\dd X}{\sqrt{f(0,X)}}. 
\end{equation}
In order to obtain $\rho(t_0,r)$ from Eq.~\eqref{densdef}, 
we need to calculate $S(t_0,r)$ and $\partial_r S(t_0,r)$ as 
functions of $r$. 
$S(t_0,r)$ can be calculated 
by numerically solving Eq.~\eqref{Eeqsol} with $t=t_0$. 
Then, we can obtain $\partial_rS(t_0,r)$ by numerically solving Eq.~\eqref{eq:drS}. 
We note that the hypersurface $t=t_0$ is a spacelike hypersurface and 
the quantity $\rho(t_0,r)$ is not a direct observable. 
Observational aspects are discussed elsewhere, and we simply use 
$\rho(t_0,r)$ to demonstrate the inhomogeneity in this paper.
Let us define the density fluctuation $\Delta_0$ as 
\begin{equation}
\Delta_0(t_0,r_{\rm lc}(z)) := \frac{ \rho(t_0,r_{\rm lc}(z))- \rho (t_0,r_{\rm lc}(10))}{\rho(t_0,r_{\rm lc}(10))}. 
\end{equation}
It should be noted that, although we describe $\Delta_0$ as a function of $z$, 
$\Delta_0(t_0,r_{\rm lc}(z))$ is defined on the spacelike surface $t=t_0$. 
The redshift $z$ is simply used to specify the radial coordinate $r$. 

\begin{figure}[htbp]
	\includegraphics[scale=0.6]{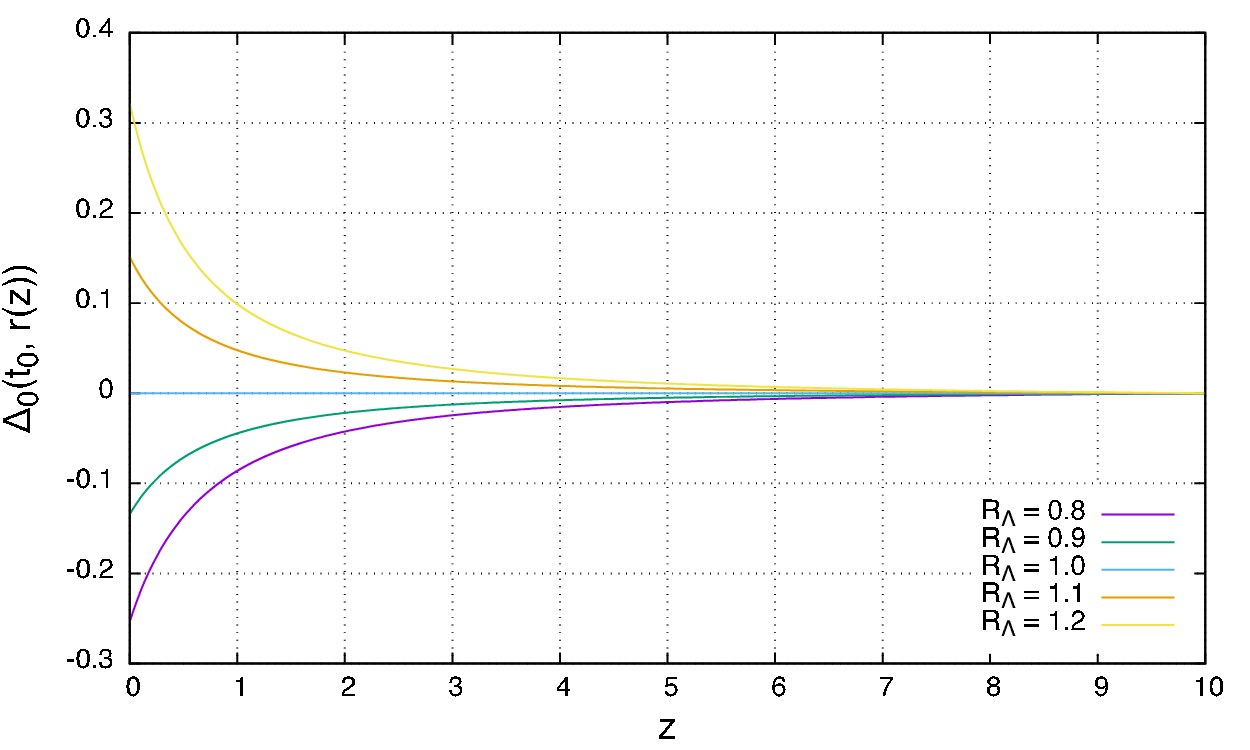}
	\includegraphics[scale=0.6]{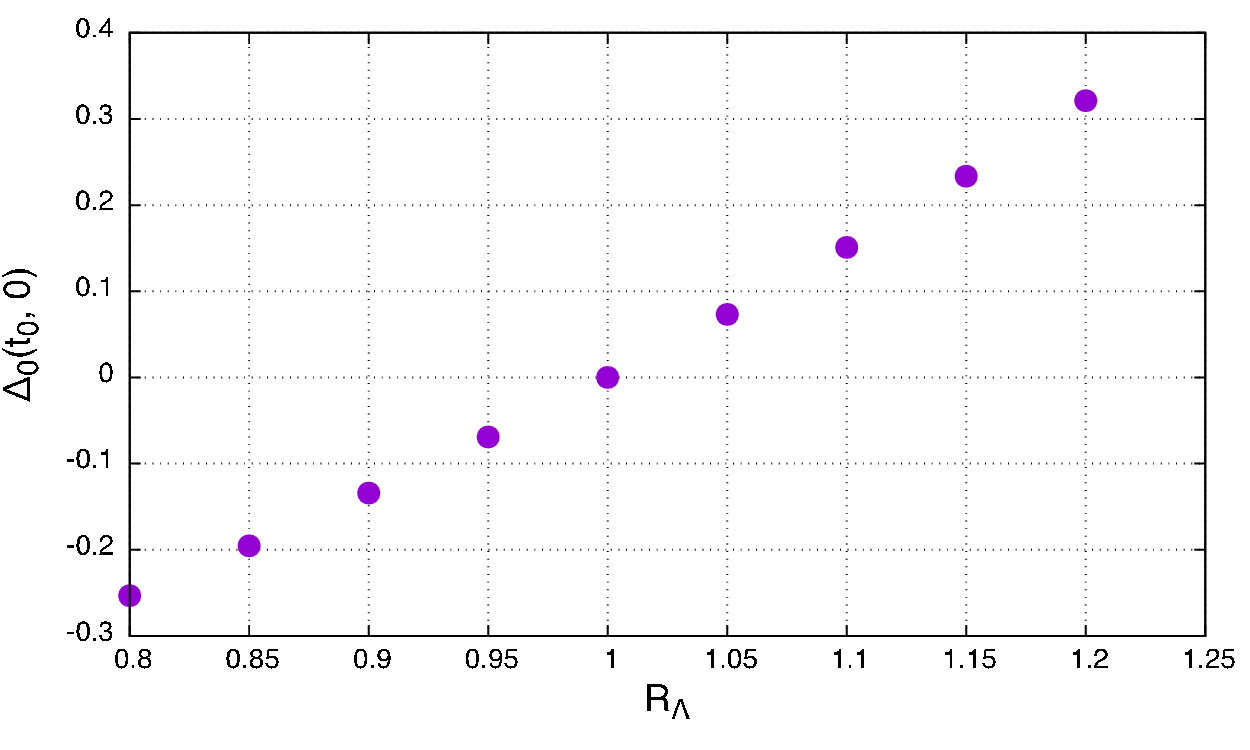}
			\caption{(left)$\Delta_0 (t_0,r_{\rm lc}(z))$ is depicted as 
			a function of $z$ for several values of $\mathcal R_\Lambda$.
			(right)The value of $\Delta_0(t_0,0)$ is depicted for several values of $\mathcal R_\Lambda$. }
			\label{delta_0}
\end{figure}

In Fig.~\ref{delta_0}, we show the density fluctuation $\Delta_0 (t_0,r_{\rm lc}(z))$ as 
a function of $z$ and the value of $\Delta_0(t_0,0)$ 
for several values of $\mathcal R_\Lambda$. 
%
As is shown in Fig.~\ref{delta_0}, 
we obtain void(over dense) structures for $\mathcal R_\Lambda < 1(>1)$ 
around the symmetry center. 
The magnitude of the density inhomogeneity is 
roughly proportional to the value of $\mathcal R_\Lambda$ and 
$\Delta_0(t_0,0)\sim-0.25$ for $\mathcal R_\Lambda=0.8$.

\section{Summary and discussion}
\label{summary and discussion}
In this paper, we have described 
technical details of the construction of the 
$\Lambda$LTB model for a given set of 
a distance-redshift relation and a value of $\Lambda$
with the bang time function being constant. 
It has been shown that we can obtain a unique $\Lambda$LTB model for 
each set of $\Lambda$ and a distance-redshift relation. 
As a demonstration, we have constructed the $\Lambda$LTB model 
whose distance-redshift relation is given by that in 
the flat $\Lambda$CDM model with the cosmological parameters 
$(\Omega^{\rm dis}_{\rm m0}, \Omega^{\rm dis}_{\rm \Lambda 0})=(0.3,0.7)$. 
As is expected from previous works, 
we obtain void type structure for a smaller value of 
$\Lambda$ compared with $3\Omega^{\rm dis}_{\rm \Lambda 0}H_0^2$. 

The method of the inverse construction 
can be a complement to the conventional method 
in which 
LTB functions($k(r)$, $M(r)$ and $t_{\rm B}(r)$ in the text) 
are directly parametrized by using 
several parameters(see, e.g. Ref.~\cite{Redlich:2014gga}). 
The models given by solving the inverse construction problem 
may be significantly different from the 
models given by the direct parametrization 
of the LTB functions. 
Therefore, it is important to combine the inverse construction method 
and analyses with observational data(see, e.g. Ref.~\cite{Sundell:2015cza}). 
We will report the CMB and local Hubble parameter analysis combined with our 
inverse construction method elsewhere\cite{withIchiki}.

\section{Acknowledgements}
We thank K. Ichiki, K. Nakao and H. Negishi for helpful comments. 
C.Y. was partially 
supported by Grant-in-Aid for Young Scientists (B) Grant Number JP16K17688
and Grant-in-Aid for Scientific Research on Innovative Areas Grant Number JP16H01097.


\end{document}